\documentclass[
reprint,
superscriptaddress,
aps,
prl,
floatfix,
]{revtex4-2}

\usepackage{natbib}
\usepackage{graphicx}
\usepackage{dcolumn}
\usepackage{bm}
\usepackage{braket}
\usepackage[version=4]{mhchem}

\usepackage[english]{babel}
\usepackage{mathtools}
\usepackage{amsmath}
\usepackage{amssymb}
\usepackage{physics}

\usepackage{xcolor}

\usepackage[caption=false]{subfig}

\usepackage[squaren]{SIunits}
\usepackage{hyperref}
\hypersetup{colorlinks=true,linkcolor=blue,citecolor=blue,urlcolor=blue}

\begin{document}

\preprint{APS/123-QED}

\title{Realization-dependent model of hopping transport in disordered media}

\author{Abel Thayil}
\affiliation{Laboratoire de Physique de la Matière Condensée, Ecole Polytechnique,\\ CNRS, Institut Polytechnique de Paris, 91120 Palaiseau, France}

\author{Marcel Filoche}
\affiliation{Institut Langevin, ESPCI Paris, Université PSL, CNRS, 75005 Paris, France}
\affiliation{Laboratoire de Physique de la Matière Condensée, Ecole Polytechnique,\\ CNRS, Institut Polytechnique de Paris, 91120 Palaiseau, France}

\date{\today}

\begin{abstract}
At low injection or low temperatures, electron transport in disordered semiconductors is dominated by phonon-assisted hopping between localized states. A very popular approach to this hopping transport is the  Miller-Abrahams model that requires a set of empirical parameters to define the hopping rates and the preferential paths between the states. We present here a transport model based on the localization landscape (LL) theory in which the location of the localized states, their energies, and the coupling between them are computed for any specific realization, accounting for its particular geometry and structure. This model unveils the transport network followed by the charge carriers that essentially consists in the geodesics of a metric deduced from the LL. The hopping rates and mobility are computed on a paradigmatic example of disordered semiconductor, and compared with the prediction from the actual solution of the Schr\"odinger equation. We explore the temperature-dependency for various disorder strengths and demonstrate the applicability of the LL theory in efficiently modeling hopping transport in disordered systems. 

\end{abstract} 

\maketitle

The classical description of electrical conduction in a semiconductor involves scattering of the electronic Bloch states on impurities or defects of the lattice. However, in nitride alloys~\cite{AleksiejunasImpactAlloyDisorderInducedLocalization2020, WeisbuchDisorderEffectsNitride2021}, perovskites~\cite{BaranowskiStaticDynamicDisorder2018a, SinghEffectThermalStructural2016} or organic semiconductors~\cite{McMahonOrganicSemiconductorsImpact2010a, TroisiChargeTransportRegimeCrystalline2006, NenashevTheoreticalToolsDescription2015}, the random arrangements of the elements in the alloy, the different inter-atomic spacings, or the random orientation of the molecules destroy the translation invariance of the crystal. In some cases, the resulting random spatial fluctuations of the local material composition are strong enough to induce localization of a large proportion of the low-energy electronic and hole states~\cite{Anderson1958}. Consequently, at low temperatures and low carrier concentrations, the charge carrier transport does not follow anymore the classical picture, but is dominated instead by phonon-assisted hopping between these localized states~\cite{MottElectronicProcessesNoncrystalline2012, ShklovskiiElectronicPropertiesDoped1984}. 

In this situation, hopping transport can be modeled as a transport process on a graph where each state (or node) is associated to an average occupation probability, and each pair of states (or edge) is associated to a transition probability or hopping rate (only close states in the nearest-neighbour model or distant in the variable range model~\cite{NenashevTheoreticalToolsDescription2015}). The dynamics of the process is then governed by a master equation that tracks down the time-evolution of the average occupation probability of each state~\cite{MillerImpurityConductionLow1960}. In the steady state, the solution to the master equation provides the equilibrium occupation probabilities and the steady state current. The input parameters to the master equation are the hopping rates which are computed by evaluating the electron-phonon interaction between each pair of states. This requires knowledge of the position and spatial extent of the wave functions for all states as well as their respective energies. When the hops are due to acoustic phonons, the atomic displacements are described as long-wavelength acoustic waves that are related to the elastic strain of the crystal, as described in the deformation potential theory~\cite{BardeenDeformationPotentialsMobilities1950}.

These quantities can be computed via \textit{ab-initio} atomistic methods~\cite{MladenovicChargeCarrierLocalization2015, ChanBridgingGapAtomic2010, MasseInitioChargecarrierMobility2016, MasseEffectsEnergyCorrelations2017} which become computationally very demanding for systems of reasonable size. Classically, this difficulty is circumvented by assuming \emph{a priori} how the localized states are distributed in space and in energy, and by providing a functional form for the hopping rates between localized states. The Miller-Abrahams (MA) model corresponds to the specific case in which electronic states are supposed to decay exponentially with one uniform localization length: the hopping rates are thus exponentially-decreasing functions of the distance between states with the same characteristic length~\cite{MillerImpurityConductionLow1960, KasuyaTheoryImpurityConduction1958a}. The free parameters of the model are fitted against experimental mobility curves. Although the MA model has been applied to a large range of organic materials~\cite{VissenbergTheoryFieldeffectMobility1998, PasveerUnifiedDescriptionChargeCarrier2005a, NenashevTheoreticalToolsDescription2015} or amorphous~\cite{Grunewald1979,  Godet2001,  Murayama2010}, one of its major drawbacks is that it relies on identical empirical parameters for all electronic states at all energies. Vukmirovi\'c \textit{et al.}~\cite{VukmirovicCarrierHoppingDisordered2010} showed via \textit{ab-initio} calculations that the exact hopping rates and mobilities can deviate significantly from the MA model, in part because the MA model does not account for the complex overlaps between the wave functions of the associated electronic states.

In this paper, we present a model of hopping transport in disordered semiconductors based on the recently developed localization landscape (LL) theory~\cite{FilocheUniversalMechanismAnderson2012} that bridges the gap between \textit{ab-initio}  atomistic calculations and empirical models such as the MA model. Our approach takes into account the structural disorder of the system and gives access to specific localization effects without significant computational cost~\cite{ArnoldEffectiveConfiningPotential2016}. The main ingredient in the LL theory is the \textit{effective potential} which not only predicts the regions of localization of the eigenstates and their corresponding energies, but also provides a fine estimate of the exponential decay of the wave functions away from their regions of existence. This enables us to compute hopping rates between localized states and consequently, the mobility of the charge carrier as a function of the underlying disordered potential. We then compare these computed mobilities with mobilities based on exact eigenstate computations for a 2D disordered potential, and analyze the dependency of the mobility against disorder strength. 

\begin{figure}
    \centering
    \includegraphics[scale = 0.05]{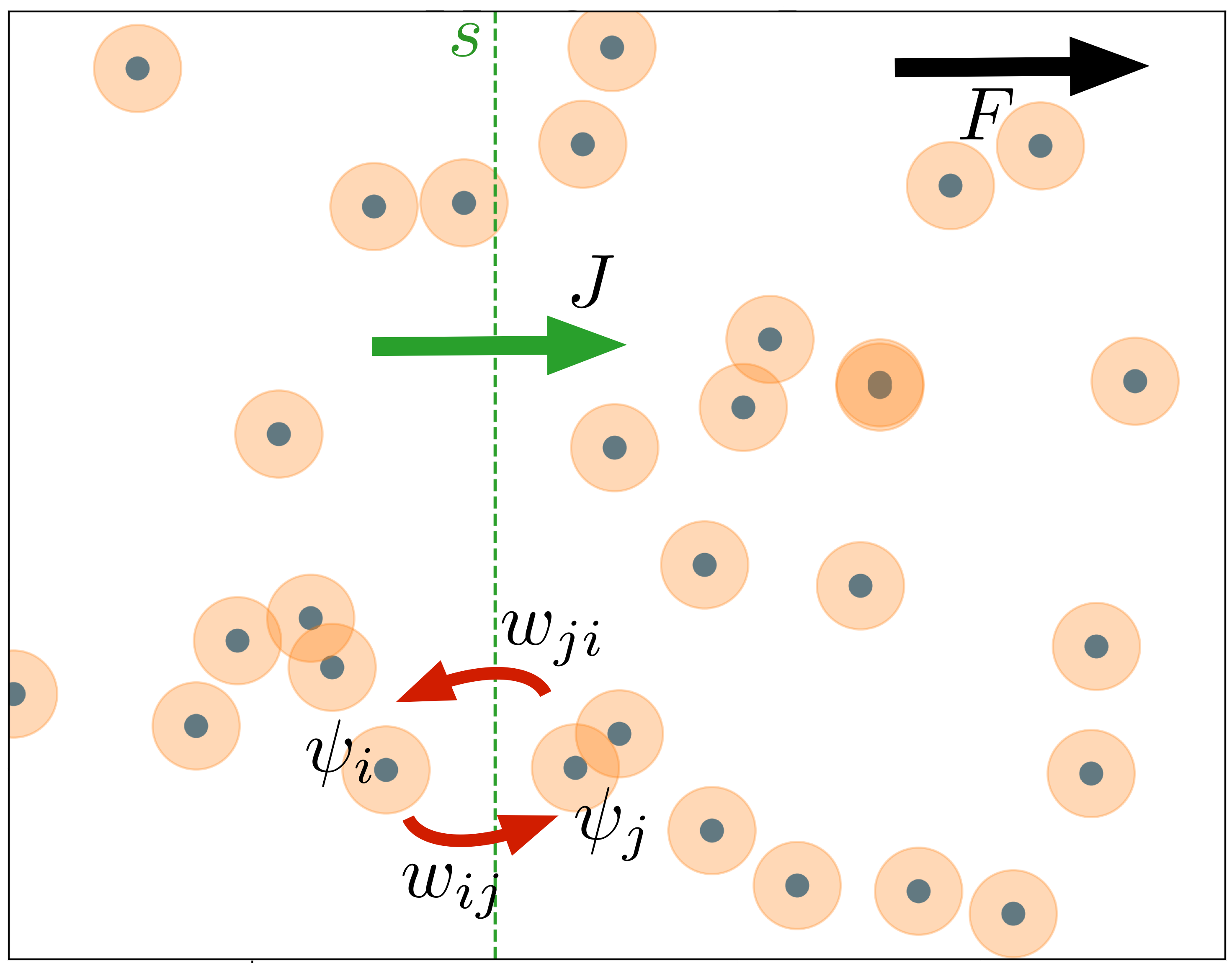}
    \caption{Schematic of hopping transport: under an applied electrostatic field $F$, electrons move by hopping (red arrows) between the localized electronic states (orange disks). The net current (green arrow) passing through the surface $S$ (green dotted line) is calculated by subtracting the net flow of charge to the left from the net flow of charge to the right.}
    \label{fig:hopping_scheme}
\end{figure}

To model hopping transport, we associate to each electronic state $i$ an average occupation probability~$f_i$. The rate of change of this occupation probability is the net sum of all the outward hops from state $i$ to any other state $j$ (with hopping rate $w_{ij}$), and inward hops from any state $j$ to the state $i$ (with hopping rate $w_{ji}$). One must also account for the fact that a carrier can only hop from an occupied to an unoccupied state. The master equation therefore reads: 
\begin{equation}
         \frac{df_i}{dt} = \sum_j \big[ -w_{ij} \, f_i (1-f_j) + w_{ji} \, f_j(1-f_i) \big] \,.
         \label{eq:master}
 \end{equation}
The hopping rates between any two states maintain detailed balance in steady state, and satisfy
\begin{equation}
\frac{w_{ij}}{w_{ji}} = \exp(\frac{E_i - E_j}{k_BT}) \,,
\end{equation}
where $E_i, E_j$ are the energies of states $i$ and $j$, respectively, $k_B$ is the Boltzmann constant, and $T$ is the temperature of the system. The presence of an electrostatic field $\vb{F}$ shifts the energy of each state, $E_i' =  E_i - e \, \vb{F} \cdot \vb{r}_i$, where $e$ is the charge of the carrier. This results in a greater number of hops in the direction of the field, and the emergence of a net current of charge carriers. The steady state current $J$ in response to the applied field $\vb{F}$ (assumed here to be along the $x$ direction) through the surface $S$ (depicted in Fig.~\ref{fig:hopping_scheme}) is 
\begin{equation}
    J = \frac{e}{L}\sum_{\substack{i \\ x_i < x_s}} \sum_{\substack{j \\ x_j > x_s}} \big[ -w_{ij}\bar{f}_i (1-\bar{f}_j) + w_{ji}\bar{f}_j(1-\bar{f}_i) \big] \,,
    \label{eq:current}
\end{equation}
where $\bar{f}_i$ is the steady state occupation probability for state $i$, and  $x_s$ is the $x$ coordinate of the surface $s$, $n$ is the carrier density and $L$ is the length of the sample. The mobility $\mu$ is then given by $\mu = J/neF$. 

The wave functions $\psi$ of the electronic states involved in the hopping process satisfy the Schr\"odinger equation,
\begin{equation}
    -\mathrm{div}\left( \frac{\hbar^{2}}{2m}\nabla\psi\right) + V\psi = E\psi \,,
    \label{eq:Schrod}
\end{equation}
where $m(\vb{r})$ is the effective mass of the charge carrier (possibly position-dependent) and $V(\vb{r})$ is the potential (in semiconductors, the conduction or valence band edge). The hopping rates between any two states $\psi_i$ and $\psi_j$ are and obtained by applying the Fermi golden rule to the electron-phonon interaction: 
\begin{align}
\label{eq:hopping_eigen}
w_{ij} = ~ \displaystyle \frac{2\pi}{\hbar} \sum_{\vb{q}} & \abs{\mel{\psi_j}{\hat{H}_{ep}}{\psi_i}}^2 \, \delta(E_j - E_i \pm E_q) \nonumber\\
& \times \Big\{ n_B(E_q,T) +\frac{1}{2} \pm \frac{1}{2}\Big\} \,,
\end{align}
where $\hat{H}_{ep}$ refers to the Hamiltonian of the electron-phonon interaction, $E_q$ is the energy of a phonon with wave vector modulus $q$, and $n_B(E_q, T)$ is the average occupation number of a phonon with energy $E_q$ at temperature $T$ (given by the Bose-Einstein statistics). For acoustic phonons treated in deformation potential theory, the hopping rate takes the value
\begin{equation}
w_{ij} = \frac{D^2q_0^3}{8\pi^2\rho_m\hbar c_s^2} \, \abs{M^{q_0}_{ij}}^2 \, \Big\{ n_B + \frac{1}{2} \pm \frac{1}{2}\Big\},
\label{eq:hopping_3d}
\end{equation}
where $D$ is the deformation potential constant, $\rho_m$ is the mass density of the material, $c_s$ is the speed of sound in the material, $q_0 = |E_j - E_i|/\hbar c_s$ and $M^{q_0}_{ij}$ is given by the following overlap integral:
\begin{equation}
M^{q_0}_{ij} = \int_{q = q_0} d\Omega_q \int d\vb{r} ~e^{-i\vb{q} \cdot \vb{r}}~\psi^{*}_i(\vb{r}) \psi_j(\vb{r}) \,.
\end{equation}

We see that the above integral depends on the spatial extent of the wave functions, and that its value is determined by the regions where the product $\abs{\psi_i(\vb{r})\psi_j(\vb{r})}$ is significant. Evaluating Eq.~\eqref{eq:hopping_3d} therefore requires knowledge of the energies, of the locations and of the spatial extents of the localized states. Solving the Schr\"odinger equation in Eq.~\eqref{eq:Schrod} to access these quantities is prohibitively expensive for large systems. In the MA model, this issue is bypassed by assuming that the localized functions exponentially decay in all directions with the same characteristic localization length~$a$, leading to the following expression for the hopping rate:
\begin{equation}
        w_{ij} = w_0 \exp( -\frac{2|\vb{r}_j - \vb{r}_i|}{a} -\frac{[E_j-E_i]_+}{k_B T}) ,
        \label{eq:MA}
\end{equation}
where $w_0$ is a typical escape frequency and $[x]_+ = \max(x,0)$. This expression corresponds to the variable-range hopping model introduced in~\cite{MottElectronicProcessesNoncrystalline2012, ShklovskiiElectronicPropertiesDoped1984}. The electronic density of states (the distribution of $E_i$) is typically assumed to be the tail of a Gaussian or of an exponential function~\cite{BaranovskiiChargeTransportDisordered2017, OelerichHowFindOut2012}. In addition, the parameters $w_0$ and $a$ need to be empirically fitted to experimental data. This drastic oversimplification of the shapes, locations, and energies of the wave functions can lead to erroneous estimates of the hopping rates, and finally of the current flowing through the system~\cite{VukmirovicCarrierHoppingDisordered2010}.

The LL theory allows us to reliably build the hopping network and assess all input parameters of the master equation without solving the Schr\"odinger equation. The LL is defined as the solution to the related Dirichlet problem, 
 \begin{equation}
    -\mathrm{div}\left( \frac{\hbar^{2}}{2m}\nabla u \right) + V u = 1.
    \label{eq:landscape}
\end{equation}
It was shown in Refs.~\cite{FilocheUniversalMechanismAnderson2012, ArnoldEffectiveConfiningPotential2016, FilocheLocalizationLandscapeTheory2017, ArnoldComputingSpectraSolving2019} that the LL~$u$ enables us to define an effective potential $V_u(\vb{r}) := 1/u(\vb{r})$ that  
\begin{itemize}

\item predicts the regions of the localization: they correspond to the basins of $V_u$~\cite{FilocheUniversalMechanismAnderson2012}. (We will see later how to precisely define these basins.)

\item provides an approximation of the fundamental eigenstate in each of these basins $B_i$~\citep{FilocheLocalizationLandscapeTheory2017}:
\begin{equation}
\psi_i(\vb{r}) \sim u(\vb{r})|_{B_i} \,
\end{equation}
up to a multiplicative constant.

\item provides an approximation $E^*_i$ of the energy $E_i$ of the local fundamental eigenstate $\psi_i$~\cite{ArnoldComputingSpectraSolving2019}:
\begin{equation}
	E_i \approx E^*_i = \left(1 + \frac{d}{4}\right) \times \min_{B_i} (V_u)  \,,
	\label{eq:energy_minimum}
\end{equation}
where $d$ is the embedding dimension of the system.

\item defines a so-called \emph{Agmon metric}~$g(\vb{r})$ and an \emph{Agmon distance} $\rho_{E}(\vb{r}_1, \vb{r}_2)$ as
\begin{align}
	g(r) = \sqrt{\frac{2m}{\hbar^2}\left[V_u(\vb{r}) - E\right]_{+}} \label{eq:agmon1} \\
	\rho_{E}(\vb{r}_1, \vb{r}_2) = \min_{\gamma(\vb{r}_1, \vb{r}_2)} \int_{\gamma} g(\vb{r}) \, ds \,, \label{eq:agmon2}
\end{align}
where the minimum is taken over all paths $\gamma$ connecting $\vb{r}_1$ to $\vb{r}_2$. This distance allows us to derive an upper bound on the exponential decay of the localized wave function via Agmon's inequality~\cite{HislopIntroductionSpectralTheory1996}:
\begin{equation}
	\psi_i(\vb{r}) ~\lesssim ~e^{-\rho_{E}(\vb{r},B_i)}
	\label{eq:agmon3}
\end{equation}
This expression can be considered as capturing very generally the quantum tunneling effect in the effective potential~$V_u$: the eigenstate~$\psi_i$ decays exponentially wherever the effective potential is larger than $E_i$ (in other words, in the barriers of $V_u$).
 
\end{itemize}

\begin{figure}
 \centering
	\subfloat[\label{fig:V}]{
		\includegraphics[width=0.48\columnwidth]{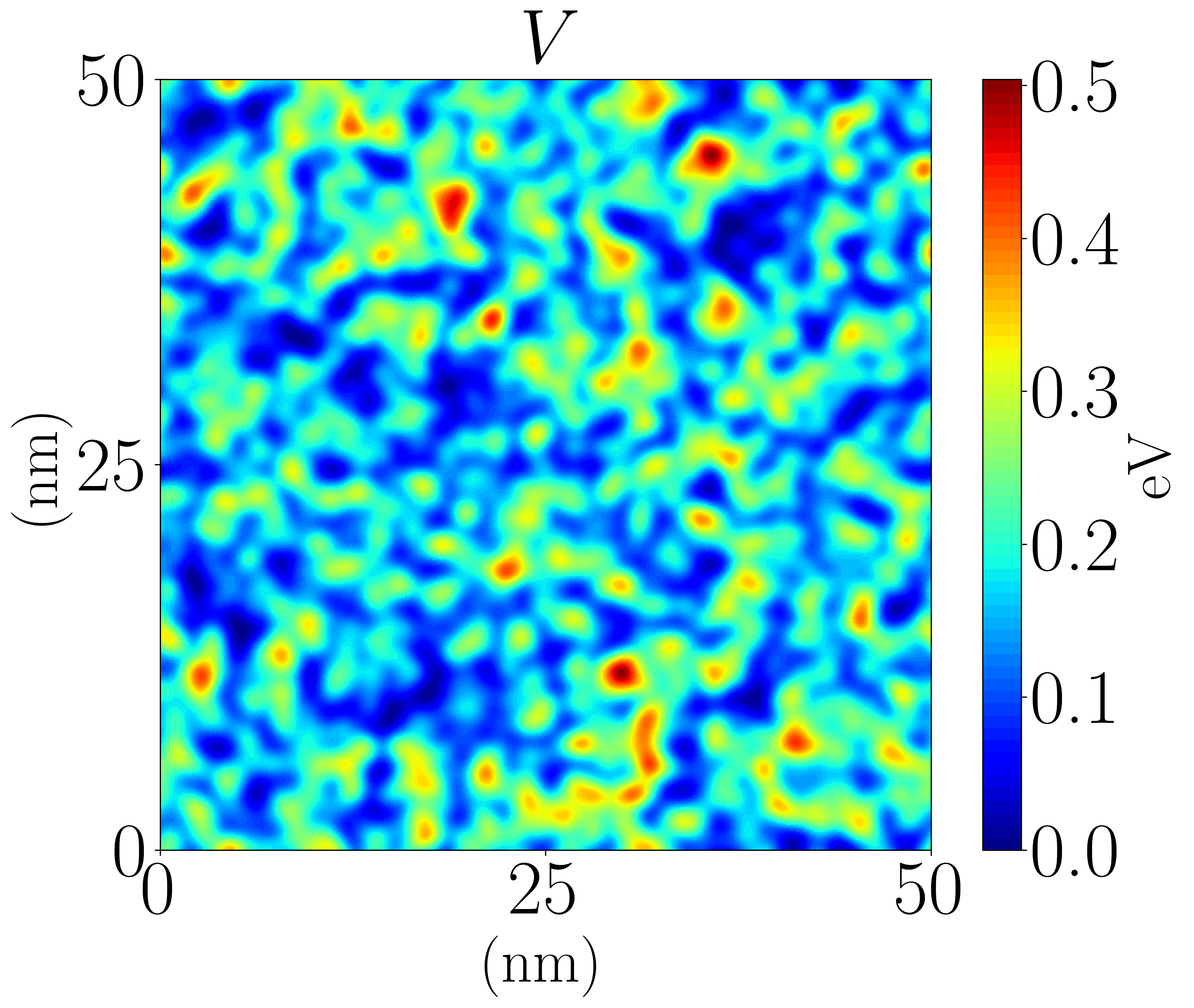}
	}\hfill
    \subfloat[\label{fig:fund_states}]{
    	\includegraphics[width=0.45\columnwidth]{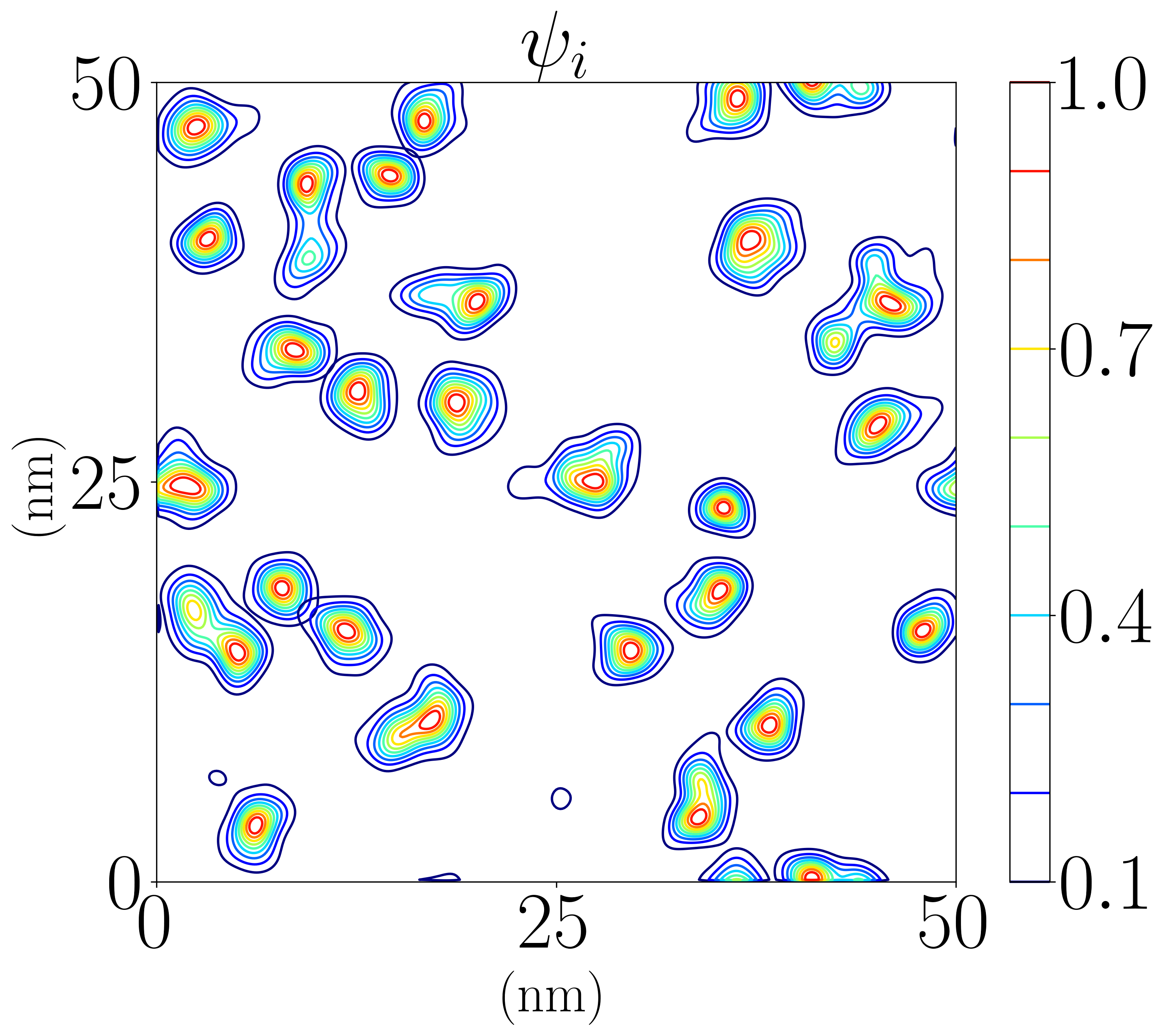}
    }

	\subfloat[\label{fig:fund_couplings}]{
		\includegraphics[width=0.46\columnwidth]{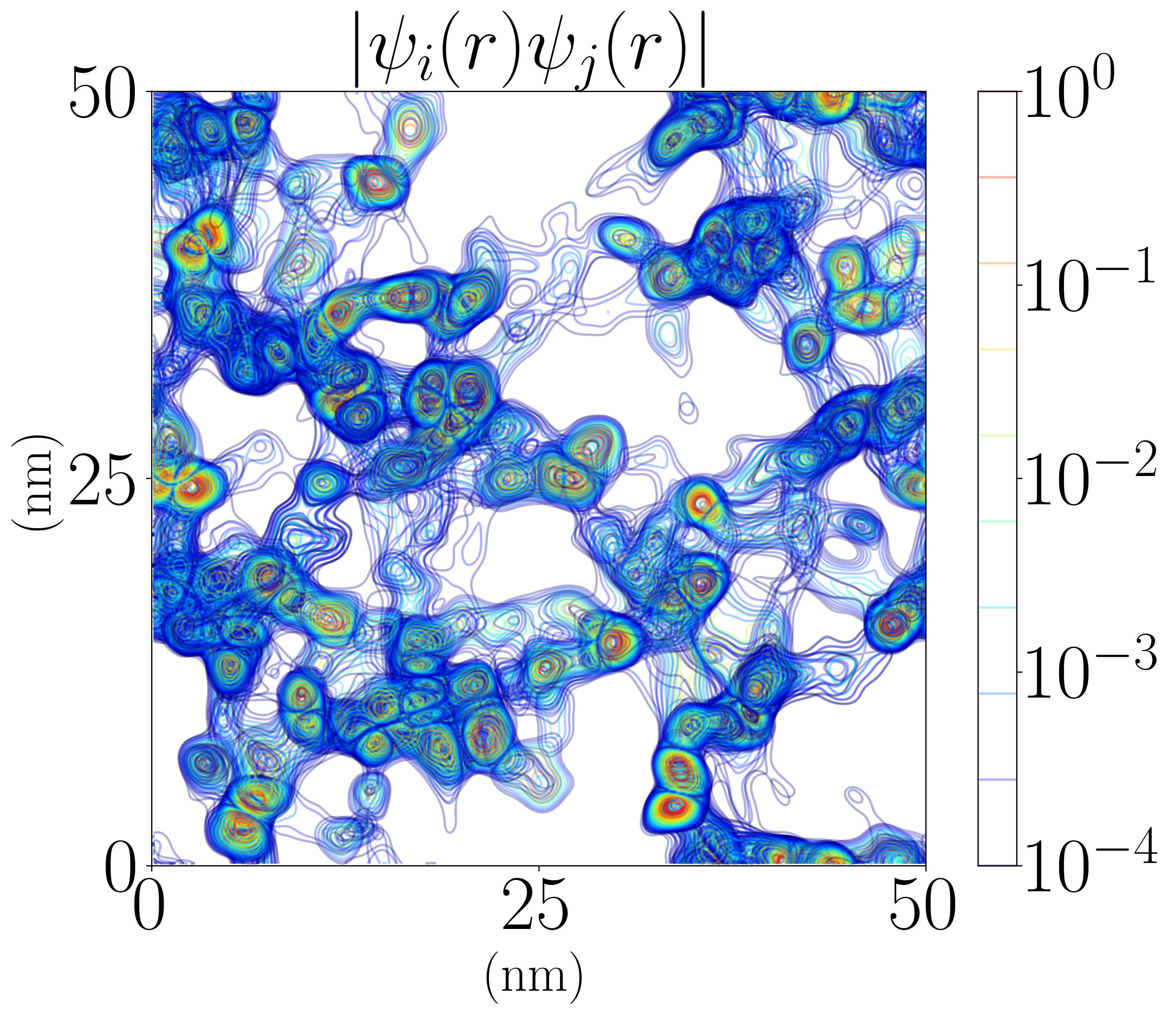}
	}\hfill
    \subfloat[\label{fig:geodesics_Vu}]{
    	\includegraphics[width=0.48\columnwidth]{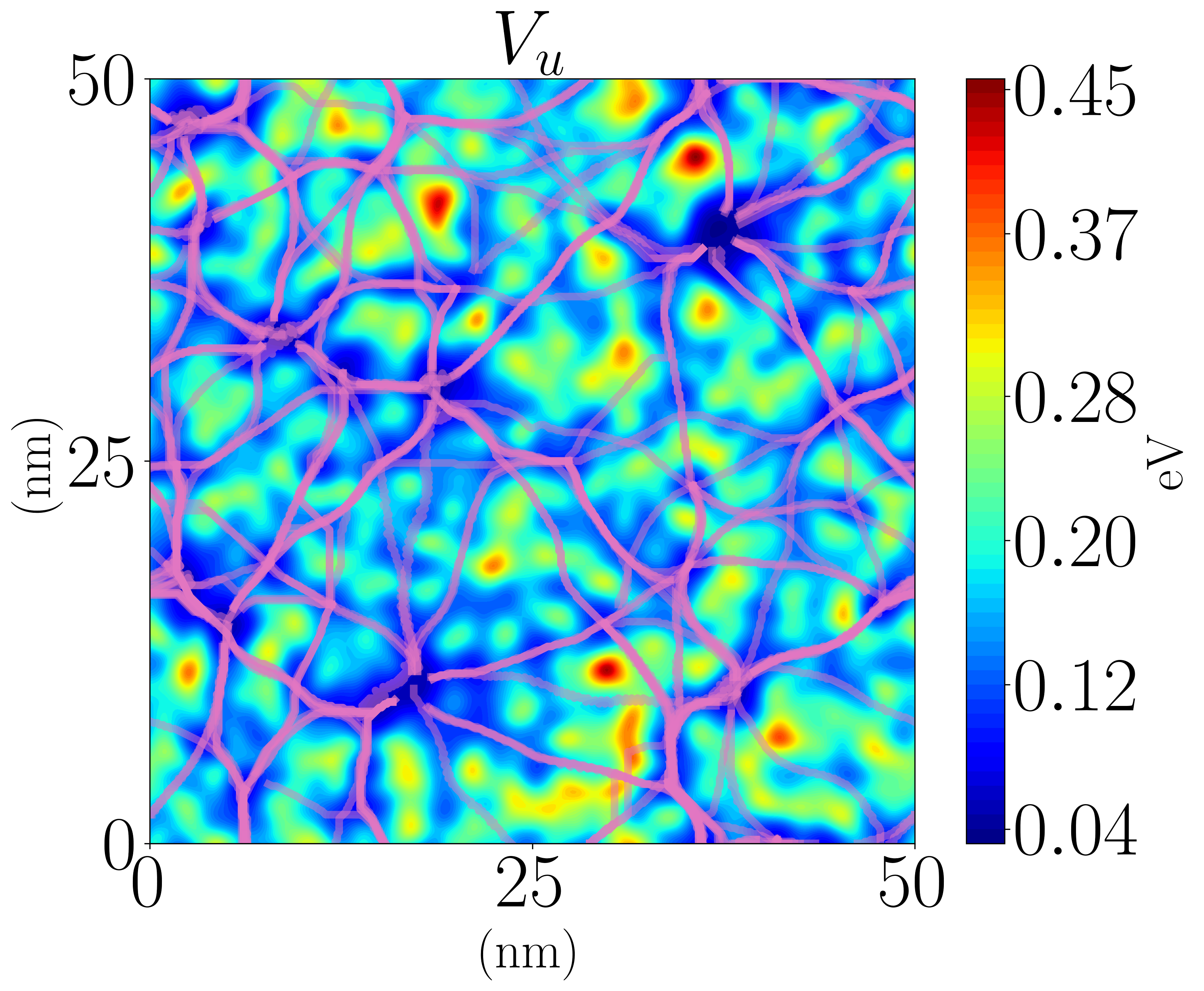}
    }
    
\caption{(a) Disordered potential with max amplitude of \unit{500}{meV}. (b) The local fundamental eigenstates of the disordered potential shown in (a). (c) Normalized pairwise overlaps $|\psi_i(\vb{r})\psi_j(\vb{r})|$ between the fundamental eigenstates. (d) The effective potential $V_u$ superimposed with the geodesics of the Agmon distance between the local minima of $V_u$ corresponding to the fundamental eigenstates (purple lines).}
\label{fig:V_states_W}
\end{figure}
    
To illustrate our approach, we study the example of a 2D disordered potential $V(\vb{r})$ depicted in Fig.~\ref{fig:V}. It can be considered as the conduction band edge of a random alloy of the form $A_x B_{1-x}$, where $A$ and $B$ are atoms placed at random on a square lattice of parameter $a=\unit{0.5}{nm}$, over a domain of size $\unit{50}{nm} \times \unit{50}{nm}$. The potential has a maximum strength of \unit{500}{meV}, and is generated by smoothing out the local composition of the alloy to define a local material, similarly as in Ref.~\cite{LiLocalizationLandscapeTheory2017a}.

To assess the accuracy of the model, we solve numerically and independently the Schr\"odinger equation with a finite element eigenvalue solver~\cite{LoggAutomatedSolutionDifferential2012, HernandezSLEPcScalableFlexible2005} to retrieve the exact eigenstates. Figure~\ref{fig:fund_states} displays the low-energy localized eigenstates. These localized eigenstates are the nodes of the hopping transport network. The connectivity of this network can be visualized by examining the product $|\psi_i(\vb{r}) \psi_j({\vb{r}})|$, as plotted in Fig.~\ref{fig:fund_couplings}. In parallel, the landscape equation (Eq. \ref{eq:landscape}) is solved using a finite element method: Figure~\ref{fig:geodesics_Vu} displays the 2D color-plot of the corresponding effective potential $V_u = 1/u$. One can see that the basins of this effective potential (in dark blue) correspond to the locations of the eigenstates.

The input parameters of the model are the energies of the states and the hopping rates between pairs of states. The former are estimated inside each basin using Eq.~\eqref{eq:energy_minimum}. The latter require first to compute the Agmon distance $\rho_{E_i}(\vb{r}, B_i)$ between each point of the domain and each basin $B_i$, see Eqs.~\eqref{eq:agmon1}-\eqref{eq:agmon2}. This can be efficiently done using a fast marching algortihm~\cite{SethianFastMarchingLevel1996, KimmelComputingGeodesicPaths1998}. We then compute  between each pair of basin minima of $V_u$ (indexed by $i$ and $j$) the geodesics of $\rho_{E_i}(\vb{r},B_i) + \rho_{E_j}(\vb{r},B_j)$. Superimposing these geodesics (in purple) over the effective potential in Fig.~\ref{fig:geodesics_Vu}, we clearly see that the network of geodesics replicates the network of pair-wise products already observed in Fig.~\ref{fig:fund_couplings}. Very generally, this approach allows us to reveal the percolation network of charge carrier trajectories giving birth to a macroscopic current, and to measure its statistical geometrical properties~\citep{AmbegaokarHoppingConductivityDisordered1971}.

This Agmon distance provides a straightforward way to reconstruct estimates $\psi^{(u)}_i$ of the eigenstates $\psi_i$:
\begin{equation}
    \psi^{(u)}_i(\vb{r}) = c_i
    \begin{dcases}
        E^*_i \, u(\vb{r}) & \text{inside}~B_i\\
        \exp(-\rho_{E^*_i}( \vb{r}, B_i)) &  \text{outside}~B_i
    \end{dcases}
\end{equation}
where $c_i$ is a normalization constant, the basin $B_i$ being defined as the connected domain around the local minimum of $1/u$ whose boundary is the level set $u(\vb{r}) = 1/E^*_i$. Through this definition, the reconstructed wave function $\psi^{(u)}_i(\vb{r})$ is everywhere continuous.

We need here to distinguish between the estimated energy $E^*_i$ of the localized state given by Eq.~\eqref{eq:energy_minimum} and the energy entering the Agmon metric in Eq.~\eqref{eq:agmon1}. Although Eq.~\eqref{eq:agmon3} holds when using the energy of the state to compute the distance $\rho_E$, a tighter bound can be obtained in practice by using a smaller value of the energy. Indeed, the Agmon distance corresponds to the path that minimizes the integral of Eq.~\eqref{eq:agmon2} while the exact value of $\psi_i$ at point~$\vb{r}$ would be obtained by a weighted sum of all possible paths through a path integral formulation, this approach being however much more computationally expensive. Since all other paths have larger distances, using the Agmon distance leads to a slight overestimation of the wave function amplitude outside its basin, hence an overestimation of the hopping rates. This effect can be compensated very simply by reducing the value of the energy entering the Agmon metric. For all potential strengths studied in our work, we have found that a value $\bar{E}_i = 1.3 \times \min_{B_i} (V_u)$ [instead of 1.5 in Eq.~\eqref{eq:energy_minimum}] works satisfactorily. This trend needs to be investigated in future studies.

The last step consists in computing the hopping rates~$w_{ij}$ using Eq.~\eqref{eq:hopping_3d}. To that end, we use material parameters similar to those of disordered alloys of InGaN: $D = \unit{8.3}{eV}$, $c_s = \unit{8 \cdot 10^3}{m \cdot s^{-1}}$ and $\rho_m = \unit{6150}{kg \cdot m^{-3}}$. A comparison of these hopping rates between exact computation, MA and LL-based models is provided in Supplementary Material. We plug these computed hopping rates into Eq.~\eqref{eq:master} and solve the master equation by a Newton-Raphson method to obtain the steady state occupation probabilities, with an initial guess for the occupation probabilities given by Fermi-Dirac statistics. For our results, we have placed the Fermi level \unit{20}{meV} below the ground state, which is a typical value due to donor states in nitride semiconductors.

\begin{figure}
\centering
\includegraphics[width=0.4\textwidth]{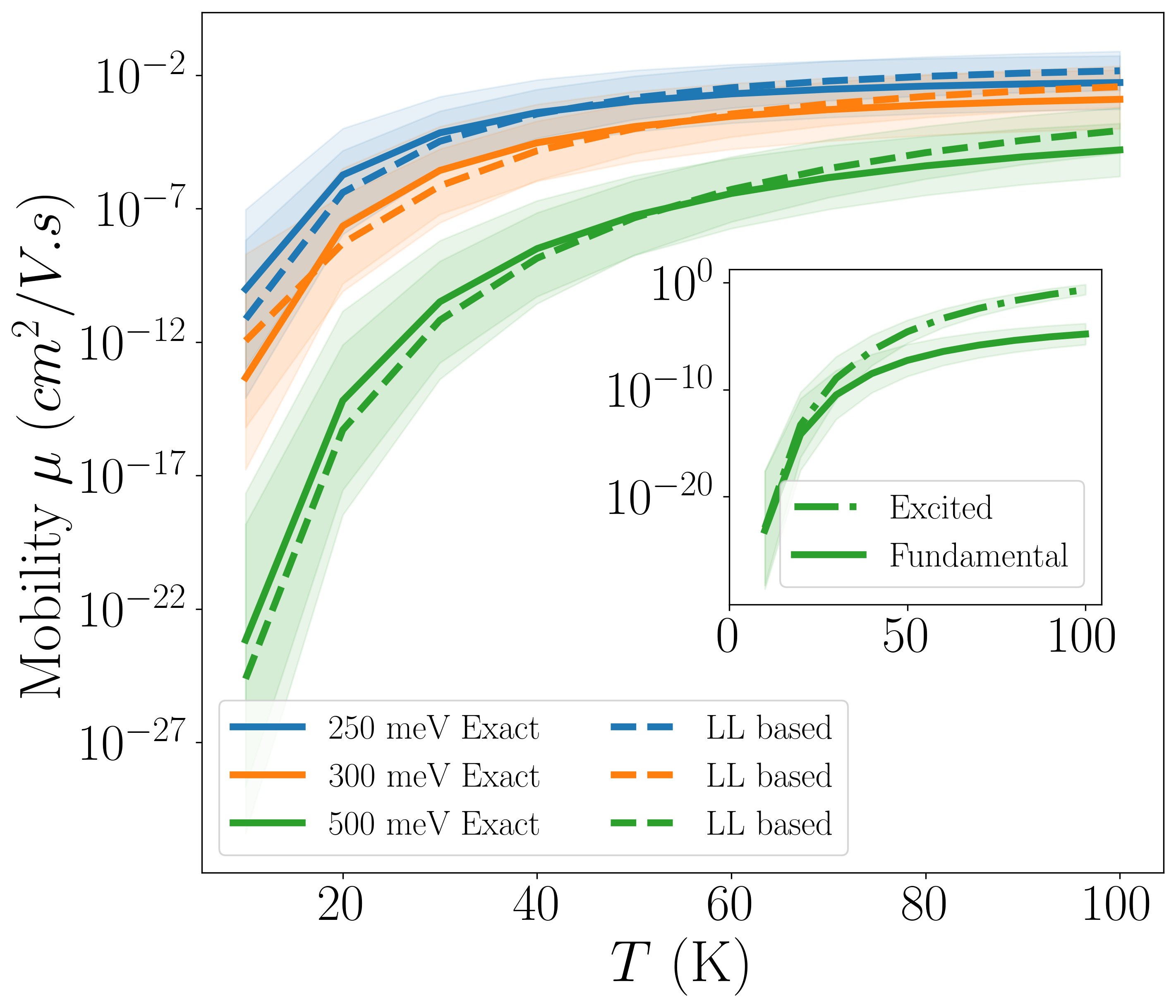}
\caption{Comparison of the hopping mobility as a function of temperature, obtained from a computation based on the fundamental eigenstates with the LL-based mobility, for three different disorder strengths averaged over 50~realizations. The shaded areas signify the standard deviation of the computed mobility. (Inset) The mobility as a function of the temperature compared for the fundamental states and both the fundamental and excited states for the disorder strength of \unit{500}{meV}. The material parameters entering Eq.~\eqref{eq:hopping_3d} are chosen as follows: $D = \unit{8.3}{eV}$, $c_s = \unit{8 \cdot 10^3}{m \cdot s^{-1}}$ and $\rho_m = \unit{6150}{kg \cdot m^{-3}}$. }
\label{fig:mob_three_potentials}
\end{figure}

Figure~\ref{fig:mob_three_potentials} compares mobility vs. temperature curves obtained by solving the master equation using parameters derived from the exact solution of the Schr\"odinger equation (solid lines), and the ones resulting from the LL-based solution of the master equation (dashed lines). The lines represent the average over 50~realizations while the shaded area corresponds to one standard deviation around the average. The LL-based computations are shown to be in very good agreement with the exact eigenstate-based computations on a wide range of temperatures, while being about 4 times faster for a 2D system of size $800 \times 800$.

One has to note that, at higher temperatures, the value of the mobility depends on the number of electronic states included in the computation. Involving more excited states increases the mobility at higher temperature (inset in Fig.~\ref{fig:mob_three_potentials}), but we chose to focus in this study on the conduction induced by the lower energy states.

In summary, the LL-based approach to hopping transport allows us to assess efficiently the carrier mobility in a highly disordered or random medium, taking into account the specific characteristics of the disorder at the nanoscale without having to solve the Schr\"odinger equation for a large number of states. This model not only encompasses naturally variable range hopping, but also provides a realization-dependent visualization of the transportation network through electron-phonon coupling between states, revealing the nature of the percolation paths followed by the charge carriers. It is therefore a very handy theoretical and practical tool for understanding the features of electronic transport at low temperatures, for testing extreme deviations of the conductivity through rare events

We are grateful to Jacques Peretti, Claude Weisbuch, Alistair Rowe, Jean-Philippe Banon and Mylene Sauty for very fruitful discussions. Both authors are supported by grants from the Simons Foundation (No 601944, M.F. and No. 1027116, M.F.).

\bibliography{hopping}

\begin{thebibliography}{36}%
\makeatletter
\providecommand \@ifxundefined [1]{%
 \@ifx{#1\undefined}
}%
\providecommand \@ifnum [1]{%
 \ifnum #1\expandafter \@firstoftwo
 \else \expandafter \@secondoftwo
 \fi
}%
\providecommand \@ifx [1]{%
 \ifx #1\expandafter \@firstoftwo
 \else \expandafter \@secondoftwo
 \fi
}%
\providecommand \natexlab [1]{#1}%
\providecommand \enquote  [1]{``#1''}%
\providecommand \bibnamefont  [1]{#1}%
\providecommand \bibfnamefont [1]{#1}%
\providecommand \citenamefont [1]{#1}%
\providecommand \href@noop [0]{\@secondoftwo}%
\providecommand \href [0]{\begingroup \@sanitize@url \@href}%
\providecommand \@href[1]{\@@startlink{#1}\@@href}%
\providecommand \@@href[1]{\endgroup#1\@@endlink}%
\providecommand \@sanitize@url [0]{\catcode `\\12\catcode `\$12\catcode
  `\&12\catcode `\#12\catcode `\^12\catcode `\_12\catcode `\%12\relax}%
\providecommand \@@startlink[1]{}%
\providecommand \@@endlink[0]{}%
\providecommand \url  [0]{\begingroup\@sanitize@url \@url }%
\providecommand \@url [1]{\endgroup\@href {#1}{\urlprefix }}%
\providecommand \urlprefix  [0]{URL }%
\providecommand \Eprint [0]{\href }%
\providecommand \doibase [0]{https://doi.org/}%
\providecommand \selectlanguage [0]{\@gobble}%
\providecommand \bibinfo  [0]{\@secondoftwo}%
\providecommand \bibfield  [0]{\@secondoftwo}%
\providecommand \translation [1]{[#1]}%
\providecommand \BibitemOpen [0]{}%
\providecommand \bibitemStop [0]{}%
\providecommand \bibitemNoStop [0]{.\EOS\space}%
\providecommand \EOS [0]{\spacefactor3000\relax}%
\providecommand \BibitemShut  [1]{\csname bibitem#1\endcsname}%
\let\auto@bib@innerbib\@empty
\bibitem [{\citenamefont {Aleksiej\ifmmode~\bar{u}\else \={u}\fi{}nas}\ \emph
  {et~al.}(2020)\citenamefont {Aleksiej\ifmmode~\bar{u}\else \={u}\fi{}nas},
  \citenamefont {Nomeika}, \citenamefont {Kravcov}, \citenamefont {Nargelas},
  \citenamefont {Kuritzky}, \citenamefont {Lynsky}, \citenamefont {Nakamura},
  \citenamefont {Weisbuch},\ and\ \citenamefont
  {Speck}}]{AleksiejunasImpactAlloyDisorderInducedLocalization2020}%
  \BibitemOpen
  \bibfield  {author} {\bibinfo {author} {\bibfnamefont {R.}~\bibnamefont
  {Aleksiej\ifmmode~\bar{u}\else \={u}\fi{}nas}}, \bibinfo {author}
  {\bibfnamefont {K.}~\bibnamefont {Nomeika}}, \bibinfo {author} {\bibfnamefont
  {O.}~\bibnamefont {Kravcov}}, \bibinfo {author} {\bibfnamefont
  {S.}~\bibnamefont {Nargelas}}, \bibinfo {author} {\bibfnamefont
  {L.}~\bibnamefont {Kuritzky}}, \bibinfo {author} {\bibfnamefont
  {C.}~\bibnamefont {Lynsky}}, \bibinfo {author} {\bibfnamefont
  {S.}~\bibnamefont {Nakamura}}, \bibinfo {author} {\bibfnamefont
  {C.}~\bibnamefont {Weisbuch}},\ and\ \bibinfo {author} {\bibfnamefont
  {J.~S.}\ \bibnamefont {Speck}},\ }\bibfield  {title} {\bibinfo {title}
  {Impact of alloy-disorder-induced localization on hole diffusion in highly
  excited $c$-plane and $m$-plane ($\mathrm{In}$,$\mathrm{Ga}$)$\mathrm{N}$
  quantum wells},\ }\href {https://doi.org/10.1103/PhysRevApplied.14.054043}
  {\bibfield  {journal} {\bibinfo  {journal} {Phys. Rev. Appl.}\ }\textbf
  {\bibinfo {volume} {14}},\ \bibinfo {pages} {054043} (\bibinfo {year}
  {2020})}\BibitemShut {NoStop}%
\bibitem [{\citenamefont {Weisbuch}\ \emph {et~al.}(2021)\citenamefont
  {Weisbuch}, \citenamefont {Nakamura}, \citenamefont {Wu},\ and\ \citenamefont
  {Speck}}]{WeisbuchDisorderEffectsNitride2021}%
  \BibitemOpen
  \bibfield  {author} {\bibinfo {author} {\bibfnamefont {C.}~\bibnamefont
  {Weisbuch}}, \bibinfo {author} {\bibfnamefont {S.}~\bibnamefont {Nakamura}},
  \bibinfo {author} {\bibfnamefont {Y.-R.}\ \bibnamefont {Wu}},\ and\ \bibinfo
  {author} {\bibfnamefont {J.~S.}\ \bibnamefont {Speck}},\ }\bibfield  {title}
  {\bibinfo {title} {Disorder effects in nitride semiconductors: Impact on
  fundamental and device properties},\ }\href
  {https://doi.org/doi:10.1515/nanoph-2020-0590} {\bibfield  {journal}
  {\bibinfo  {journal} {Nanophotonics}\ }\textbf {\bibinfo {volume} {10}},\
  \bibinfo {pages} {3} (\bibinfo {year} {2021})}\BibitemShut {NoStop}%
\bibitem [{\citenamefont {Baranowski}\ \emph {et~al.}(2018)\citenamefont
  {Baranowski}, \citenamefont {Urban}, \citenamefont {Zhang}, \citenamefont
  {Surrente}, \citenamefont {Maude}, \citenamefont {{Andaji-Garmaroudi}},
  \citenamefont {Stranks},\ and\ \citenamefont
  {Plochocka}}]{BaranowskiStaticDynamicDisorder2018a}%
  \BibitemOpen
  \bibfield  {author} {\bibinfo {author} {\bibfnamefont {M.}~\bibnamefont
  {Baranowski}}, \bibinfo {author} {\bibfnamefont {J.~M.}\ \bibnamefont
  {Urban}}, \bibinfo {author} {\bibfnamefont {N.}~\bibnamefont {Zhang}},
  \bibinfo {author} {\bibfnamefont {A.}~\bibnamefont {Surrente}}, \bibinfo
  {author} {\bibfnamefont {D.~K.}\ \bibnamefont {Maude}}, \bibinfo {author}
  {\bibfnamefont {Z.}~\bibnamefont {{Andaji-Garmaroudi}}}, \bibinfo {author}
  {\bibfnamefont {S.~D.}\ \bibnamefont {Stranks}},\ and\ \bibinfo {author}
  {\bibfnamefont {P.}~\bibnamefont {Plochocka}},\ }\bibfield  {title} {\bibinfo
  {title} {Static and {{Dynamic Disorder}} in {{Triple}}-{{Cation Hybrid
  Perovskites}}},\ }\href {https://doi.org/10.1021/acs.jpcc.8b05222} {\bibfield
   {journal} {\bibinfo  {journal} {J. Phys. Chem. C}\ }\textbf {\bibinfo
  {volume} {122}},\ \bibinfo {pages} {17473} (\bibinfo {year}
  {2018})}\BibitemShut {NoStop}%
\bibitem [{\citenamefont {Singh}\ \emph {et~al.}(2016)\citenamefont {Singh},
  \citenamefont {Li}, \citenamefont {Panzer}, \citenamefont {Narasimhan},
  \citenamefont {Graeser}, \citenamefont {Gujar}, \citenamefont {K{\"o}hler},
  \citenamefont {Thelakkat}, \citenamefont {Huettner},\ and\ \citenamefont
  {Kabra}}]{SinghEffectThermalStructural2016}%
  \BibitemOpen
  \bibfield  {author} {\bibinfo {author} {\bibfnamefont {S.}~\bibnamefont
  {Singh}}, \bibinfo {author} {\bibfnamefont {C.}~\bibnamefont {Li}}, \bibinfo
  {author} {\bibfnamefont {F.}~\bibnamefont {Panzer}}, \bibinfo {author}
  {\bibfnamefont {K.~L.}\ \bibnamefont {Narasimhan}}, \bibinfo {author}
  {\bibfnamefont {A.}~\bibnamefont {Graeser}}, \bibinfo {author} {\bibfnamefont
  {T.~P.}\ \bibnamefont {Gujar}}, \bibinfo {author} {\bibfnamefont
  {A.}~\bibnamefont {K{\"o}hler}}, \bibinfo {author} {\bibfnamefont
  {M.}~\bibnamefont {Thelakkat}}, \bibinfo {author} {\bibfnamefont
  {S.}~\bibnamefont {Huettner}},\ and\ \bibinfo {author} {\bibfnamefont
  {D.}~\bibnamefont {Kabra}},\ }\bibfield  {title} {\bibinfo {title} {Effect of
  thermal and structural disorder on the electronic structure of hybrid
  perovskite semiconductor {CH$_3$NH$_3$PbI$_3$}},\ }\href
  {https://doi.org/10.1021/acs.jpclett.6b01207} {\bibfield  {journal} {\bibinfo
   {journal} {J. Phys. Chem. Lett.}\ }\textbf {\bibinfo {volume} {7}},\
  \bibinfo {pages} {3014} (\bibinfo {year} {2016})}\BibitemShut {NoStop}%
\bibitem [{\citenamefont {McMahon}\ and\ \citenamefont
  {Troisi}(2010)}]{McMahonOrganicSemiconductorsImpact2010a}%
  \BibitemOpen
  \bibfield  {author} {\bibinfo {author} {\bibfnamefont {D.~P.}\ \bibnamefont
  {McMahon}}\ and\ \bibinfo {author} {\bibfnamefont {A.}~\bibnamefont
  {Troisi}},\ }\bibfield  {title} {\bibinfo {title} {Organic semiconductors:
  Impact of disorder at different timescales},\ }\href
  {https://doi.org/https://doi.org/10.1002/cphc.201000182} {\bibfield
  {journal} {\bibinfo  {journal} {ChemPhysChem}\ }\textbf {\bibinfo {volume}
  {11}},\ \bibinfo {pages} {2067} (\bibinfo {year} {2010})}\BibitemShut
  {NoStop}%
\bibitem [{\citenamefont {Troisi}\ and\ \citenamefont
  {Orlandi}(2006)}]{TroisiChargeTransportRegimeCrystalline2006}%
  \BibitemOpen
  \bibfield  {author} {\bibinfo {author} {\bibfnamefont {A.}~\bibnamefont
  {Troisi}}\ and\ \bibinfo {author} {\bibfnamefont {G.}~\bibnamefont
  {Orlandi}},\ }\bibfield  {title} {\bibinfo {title} {Charge-transport regime
  of crystalline organic semiconductors: Diffusion limited by thermal
  off-diagonal electronic disorder},\ }\href
  {https://doi.org/10.1103/PhysRevLett.96.086601} {\bibfield  {journal}
  {\bibinfo  {journal} {Phys. Rev. Lett.}\ }\textbf {\bibinfo {volume} {96}},\
  \bibinfo {pages} {086601} (\bibinfo {year} {2006})}\BibitemShut {NoStop}%
\bibitem [{\citenamefont {Nenashev}\ \emph {et~al.}(2015)\citenamefont
  {Nenashev}, \citenamefont {Oelerich},\ and\ \citenamefont
  {Baranovskii}}]{NenashevTheoreticalToolsDescription2015}%
  \BibitemOpen
  \bibfield  {author} {\bibinfo {author} {\bibfnamefont {A.~V.}\ \bibnamefont
  {Nenashev}}, \bibinfo {author} {\bibfnamefont {J.~O.}\ \bibnamefont
  {Oelerich}},\ and\ \bibinfo {author} {\bibfnamefont {S.~D.}\ \bibnamefont
  {Baranovskii}},\ }\bibfield  {title} {\bibinfo {title} {Theoretical tools for
  the description of charge transport in disordered organic semiconductors},\
  }\href {https://doi.org/10.1088/0953-8984/27/9/093201} {\bibfield  {journal}
  {\bibinfo  {journal} {J. Phys.: Condens. Matter}\ }\textbf {\bibinfo {volume}
  {27}},\ \bibinfo {pages} {093201} (\bibinfo {year} {2015})}\BibitemShut
  {NoStop}%
\bibitem [{\citenamefont {Anderson}(1958)}]{Anderson1958}%
  \BibitemOpen
  \bibfield  {author} {\bibinfo {author} {\bibfnamefont {P.~W.}\ \bibnamefont
  {Anderson}},\ }\bibfield  {title} {\bibinfo {title} {Absence of diffusion in
  certain random lattices},\ }\href {https://doi.org/10.1103/PhysRev.109.1492}
  {\bibfield  {journal} {\bibinfo  {journal} {Phys. Rev.}\ }\textbf {\bibinfo
  {volume} {109}},\ \bibinfo {pages} {1492} (\bibinfo {year}
  {1958})}\BibitemShut {NoStop}%
\bibitem [{\citenamefont {Mott}\ and\ \citenamefont
  {Davis}(2012)}]{MottElectronicProcessesNoncrystalline2012}%
  \BibitemOpen
  \bibfield  {author} {\bibinfo {author} {\bibfnamefont {N.~F.}\ \bibnamefont
  {Mott}}\ and\ \bibinfo {author} {\bibfnamefont {E.~A.}\ \bibnamefont
  {Davis}},\ }\href@noop {} {\emph {\bibinfo {title} {Electronic Processes in
  Non-Crystalline Materials}}},\ \bibinfo {edition} {2nd}\ ed.,\ The
  International Series of Monographs on Physics\ (\bibinfo  {publisher}
  {{Oxford Univ. Press}},\ \bibinfo {year} {2012})\BibitemShut {NoStop}%
\bibitem [{\citenamefont {Shklovskii}\ and\ \citenamefont
  {Efros}(1984)}]{ShklovskiiElectronicPropertiesDoped1984}%
  \BibitemOpen
  \bibfield  {author} {\bibinfo {author} {\bibfnamefont {B.~I.}\ \bibnamefont
  {Shklovskii}}\ and\ \bibinfo {author} {\bibfnamefont {A.~L.}\ \bibnamefont
  {Efros}},\ }\href {https://doi.org/10.1007/978-3-662-02403-4} {\emph
  {\bibinfo {title} {Electronic {{Properties}} of {{Doped Semiconductors}}}}},\
  \bibinfo {series} {Springer {{Series}} in {{Solid-State Sciences}}},
  Vol.~\bibinfo {volume} {45}\ (\bibinfo  {publisher} {{Springer Berlin
  Heidelberg}},\ \bibinfo {year} {1984})\BibitemShut {NoStop}%
\bibitem [{\citenamefont {Miller}\ and\ \citenamefont
  {Abrahams}(1960)}]{MillerImpurityConductionLow1960}%
  \BibitemOpen
  \bibfield  {author} {\bibinfo {author} {\bibfnamefont {A.}~\bibnamefont
  {Miller}}\ and\ \bibinfo {author} {\bibfnamefont {E.}~\bibnamefont
  {Abrahams}},\ }\bibfield  {title} {\bibinfo {title} {Impurity conduction at
  low concentrations},\ }\href {https://doi.org/10.1103/PhysRev.120.745}
  {\bibfield  {journal} {\bibinfo  {journal} {Phys. Rev.}\ }\textbf {\bibinfo
  {volume} {120}},\ \bibinfo {pages} {745} (\bibinfo {year}
  {1960})}\BibitemShut {NoStop}%
\bibitem [{\citenamefont {Bardeen}\ and\ \citenamefont
  {Shockley}(1950)}]{BardeenDeformationPotentialsMobilities1950}%
  \BibitemOpen
  \bibfield  {author} {\bibinfo {author} {\bibfnamefont {J.}~\bibnamefont
  {Bardeen}}\ and\ \bibinfo {author} {\bibfnamefont {W.}~\bibnamefont
  {Shockley}},\ }\bibfield  {title} {\bibinfo {title} {Deformation potentials
  and mobilities in non-polar crystals},\ }\href
  {https://doi.org/10.1103/PhysRev.80.72} {\bibfield  {journal} {\bibinfo
  {journal} {Phys. Rev.}\ }\textbf {\bibinfo {volume} {80}},\ \bibinfo {pages}
  {72} (\bibinfo {year} {1950})}\BibitemShut {NoStop}%
\bibitem [{\citenamefont {Mladenovi{\'c}}\ and\ \citenamefont
  {Vukmirovi{\'c}}(2015)}]{MladenovicChargeCarrierLocalization2015}%
  \BibitemOpen
  \bibfield  {author} {\bibinfo {author} {\bibfnamefont {M.}~\bibnamefont
  {Mladenovi{\'c}}}\ and\ \bibinfo {author} {\bibfnamefont {N.}~\bibnamefont
  {Vukmirovi{\'c}}},\ }\bibfield  {title} {\bibinfo {title} {Charge carrier
  localization and transport in organic semiconductors: Insights from atomistic
  multiscale simulations},\ }\href
  {https://doi.org/https://doi.org/10.1002/adfm.201402435} {\bibfield
  {journal} {\bibinfo  {journal} {Adv. Funct. Mater.}\ }\textbf {\bibinfo
  {volume} {25}},\ \bibinfo {pages} {1915} (\bibinfo {year}
  {2015})}\BibitemShut {NoStop}%
\bibitem [{\citenamefont {Chan}\ \emph {et~al.}(2010)\citenamefont {Chan},
  \citenamefont {Liu},\ and\ \citenamefont
  {Zunger}}]{ChanBridgingGapAtomic2010}%
  \BibitemOpen
  \bibfield  {author} {\bibinfo {author} {\bibfnamefont {J.~A.}\ \bibnamefont
  {Chan}}, \bibinfo {author} {\bibfnamefont {J.~Z.}\ \bibnamefont {Liu}},\ and\
  \bibinfo {author} {\bibfnamefont {A.}~\bibnamefont {Zunger}},\ }\bibfield
  {title} {\bibinfo {title} {Bridging the gap between atomic microstructure and
  electronic properties of alloys: The case of ({In},{Ga}){N}},\ }\href
  {https://doi.org/10.1103/PhysRevB.82.045112} {\bibfield  {journal} {\bibinfo
  {journal} {Phys. Rev. B}\ }\textbf {\bibinfo {volume} {82}},\ \bibinfo
  {pages} {045112} (\bibinfo {year} {2010})}\BibitemShut {NoStop}%
\bibitem [{\citenamefont {Mass\'e}\ \emph {et~al.}(2016)\citenamefont
  {Mass\'e}, \citenamefont {Friederich}, \citenamefont {Symalla}, \citenamefont
  {Liu}, \citenamefont {Nitsche}, \citenamefont {Coehoorn}, \citenamefont
  {Wenzel},\ and\ \citenamefont
  {Bobbert}}]{MasseInitioChargecarrierMobility2016}%
  \BibitemOpen
  \bibfield  {author} {\bibinfo {author} {\bibfnamefont {A.}~\bibnamefont
  {Mass\'e}}, \bibinfo {author} {\bibfnamefont {P.}~\bibnamefont {Friederich}},
  \bibinfo {author} {\bibfnamefont {F.}~\bibnamefont {Symalla}}, \bibinfo
  {author} {\bibfnamefont {F.}~\bibnamefont {Liu}}, \bibinfo {author}
  {\bibfnamefont {R.}~\bibnamefont {Nitsche}}, \bibinfo {author} {\bibfnamefont
  {R.}~\bibnamefont {Coehoorn}}, \bibinfo {author} {\bibfnamefont
  {W.}~\bibnamefont {Wenzel}},\ and\ \bibinfo {author} {\bibfnamefont {P.~A.}\
  \bibnamefont {Bobbert}},\ }\bibfield  {title} {\bibinfo {title} {Ab initio
  charge-carrier mobility model for amorphous molecular semiconductors},\
  }\href {https://doi.org/10.1103/PhysRevB.93.195209} {\bibfield  {journal}
  {\bibinfo  {journal} {Phys. Rev. B}\ }\textbf {\bibinfo {volume} {93}},\
  \bibinfo {pages} {195209} (\bibinfo {year} {2016})}\BibitemShut {NoStop}%
\bibitem [{\citenamefont {Mass\'e}\ \emph {et~al.}(2017)\citenamefont
  {Mass\'e}, \citenamefont {Friederich}, \citenamefont {Symalla}, \citenamefont
  {Liu}, \citenamefont {Meded}, \citenamefont {Coehoorn}, \citenamefont
  {Wenzel},\ and\ \citenamefont
  {Bobbert}}]{MasseEffectsEnergyCorrelations2017}%
  \BibitemOpen
  \bibfield  {author} {\bibinfo {author} {\bibfnamefont {A.}~\bibnamefont
  {Mass\'e}}, \bibinfo {author} {\bibfnamefont {P.}~\bibnamefont {Friederich}},
  \bibinfo {author} {\bibfnamefont {F.}~\bibnamefont {Symalla}}, \bibinfo
  {author} {\bibfnamefont {F.}~\bibnamefont {Liu}}, \bibinfo {author}
  {\bibfnamefont {V.}~\bibnamefont {Meded}}, \bibinfo {author} {\bibfnamefont
  {R.}~\bibnamefont {Coehoorn}}, \bibinfo {author} {\bibfnamefont
  {W.}~\bibnamefont {Wenzel}},\ and\ \bibinfo {author} {\bibfnamefont {P.~A.}\
  \bibnamefont {Bobbert}},\ }\bibfield  {title} {\bibinfo {title} {Effects of
  energy correlations and superexchange on charge transport and exciton
  formation in amorphous molecular semiconductors: An ab initio study},\ }\href
  {https://doi.org/10.1103/PhysRevB.95.115204} {\bibfield  {journal} {\bibinfo
  {journal} {Phys. Rev. B}\ }\textbf {\bibinfo {volume} {95}},\ \bibinfo
  {pages} {115204} (\bibinfo {year} {2017})}\BibitemShut {NoStop}%
\bibitem [{\citenamefont {Kasuya}\ and\ \citenamefont
  {Koide}(1958)}]{KasuyaTheoryImpurityConduction1958a}%
  \BibitemOpen
  \bibfield  {author} {\bibinfo {author} {\bibfnamefont {T.}~\bibnamefont
  {Kasuya}}\ and\ \bibinfo {author} {\bibfnamefont {S.}~\bibnamefont {Koide}},\
  }\bibfield  {title} {\bibinfo {title} {{A Theory of Impurity Conduction.
  II}},\ }\href {https://doi.org/10.1143/JPSJ.13.1287} {\bibfield  {journal}
  {\bibinfo  {journal} {J. Phys. Soc. Japan}\ }\textbf {\bibinfo {volume}
  {13}},\ \bibinfo {pages} {1287} (\bibinfo {year} {1958})}\BibitemShut
  {NoStop}%
\bibitem [{\citenamefont {Vissenberg}\ and\ \citenamefont
  {Matters}(1998)}]{VissenbergTheoryFieldeffectMobility1998}%
  \BibitemOpen
  \bibfield  {author} {\bibinfo {author} {\bibfnamefont {M.~C. J.~M.}\
  \bibnamefont {Vissenberg}}\ and\ \bibinfo {author} {\bibfnamefont
  {M.}~\bibnamefont {Matters}},\ }\bibfield  {title} {\bibinfo {title} {Theory
  of the field-effect mobility in amorphous organic transistors},\ }\href
  {https://doi.org/10.1103/PhysRevB.57.12964} {\bibfield  {journal} {\bibinfo
  {journal} {Phys. Rev. B}\ }\textbf {\bibinfo {volume} {57}},\ \bibinfo
  {pages} {12964} (\bibinfo {year} {1998})}\BibitemShut {NoStop}%
\bibitem [{\citenamefont {Pasveer}\ \emph {et~al.}(2005)\citenamefont
  {Pasveer}, \citenamefont {Cottaar}, \citenamefont {Tanase}, \citenamefont
  {Coehoorn}, \citenamefont {Bobbert}, \citenamefont {Blom}, \citenamefont
  {de~Leeuw},\ and\ \citenamefont
  {Michels}}]{PasveerUnifiedDescriptionChargeCarrier2005a}%
  \BibitemOpen
  \bibfield  {author} {\bibinfo {author} {\bibfnamefont {W.~F.}\ \bibnamefont
  {Pasveer}}, \bibinfo {author} {\bibfnamefont {J.}~\bibnamefont {Cottaar}},
  \bibinfo {author} {\bibfnamefont {C.}~\bibnamefont {Tanase}}, \bibinfo
  {author} {\bibfnamefont {R.}~\bibnamefont {Coehoorn}}, \bibinfo {author}
  {\bibfnamefont {P.~A.}\ \bibnamefont {Bobbert}}, \bibinfo {author}
  {\bibfnamefont {P.~W.~M.}\ \bibnamefont {Blom}}, \bibinfo {author}
  {\bibfnamefont {D.~M.}\ \bibnamefont {de~Leeuw}},\ and\ \bibinfo {author}
  {\bibfnamefont {M.~A.~J.}\ \bibnamefont {Michels}},\ }\bibfield  {title}
  {\bibinfo {title} {Unified description of charge-carrier mobilities in
  disordered semiconducting polymers},\ }\href
  {https://doi.org/10.1103/PhysRevLett.94.206601} {\bibfield  {journal}
  {\bibinfo  {journal} {Phys. Rev. Lett.}\ }\textbf {\bibinfo {volume} {94}},\
  \bibinfo {pages} {206601} (\bibinfo {year} {2005})}\BibitemShut {NoStop}%
\bibitem [{\citenamefont {Gr{\"u}newald}\ and\ \citenamefont
  {Thomas}(1979)}]{Grunewald1979}%
  \BibitemOpen
  \bibfield  {author} {\bibinfo {author} {\bibfnamefont {M.}~\bibnamefont
  {Gr{\"u}newald}}\ and\ \bibinfo {author} {\bibfnamefont {P.}~\bibnamefont
  {Thomas}},\ }\bibfield  {title} {\bibinfo {title} {A hopping model for
  activated charge transport in amorphous silicon},\ }\href
  {https://doi.org/https://doi.org/10.1002/pssb.2220940113} {\bibfield
  {journal} {\bibinfo  {journal} {Phys. Status Solidi B}\ }\textbf {\bibinfo
  {volume} {94}},\ \bibinfo {pages} {125} (\bibinfo {year} {1979})}\BibitemShut
  {NoStop}%
\bibitem [{\citenamefont {Godet}(2001)}]{Godet2001}%
  \BibitemOpen
  \bibfield  {author} {\bibinfo {author} {\bibfnamefont {C.}~\bibnamefont
  {Godet}},\ }\bibfield  {title} {\bibinfo {title} {Hopping model for charge
  transport in amorphous carbon},\ }\href
  {https://doi.org/10.1080/13642810108216536} {\bibfield  {journal} {\bibinfo
  {journal} {Philos. mag. B}\ }\textbf {\bibinfo {volume} {81}},\ \bibinfo
  {pages} {205} (\bibinfo {year} {2001})}\BibitemShut {NoStop}%
\bibitem [{\citenamefont {Murayama}\ \emph {et~al.}(2010)\citenamefont
  {Murayama}, \citenamefont {Nomura},\ and\ \citenamefont
  {Fujisaki}}]{Murayama2010}%
  \BibitemOpen
  \bibfield  {author} {\bibinfo {author} {\bibfnamefont {K.}~\bibnamefont
  {Murayama}}, \bibinfo {author} {\bibfnamefont {Y.}~\bibnamefont {Nomura}},\
  and\ \bibinfo {author} {\bibfnamefont {T.}~\bibnamefont {Fujisaki}},\
  }\bibfield  {title} {\bibinfo {title} {Hopping transport at localized band
  tail states in amorphous hydrogenated silicon},\ }\href
  {https://doi.org/https://doi.org/10.1002/pssa.200982744} {\bibfield
  {journal} {\bibinfo  {journal} {Phys. Status Solidi A}\ }\textbf {\bibinfo
  {volume} {207}},\ \bibinfo {pages} {561} (\bibinfo {year}
  {2010})}\BibitemShut {NoStop}%
\bibitem [{\citenamefont {Vukmirovi{\'c}}\ and\ \citenamefont
  {Wang}(2010)}]{VukmirovicCarrierHoppingDisordered2010}%
  \BibitemOpen
  \bibfield  {author} {\bibinfo {author} {\bibfnamefont {N.}~\bibnamefont
  {Vukmirovi{\'c}}}\ and\ \bibinfo {author} {\bibfnamefont {L.-W.}\
  \bibnamefont {Wang}},\ }\bibfield  {title} {\bibinfo {title} {{Carrier
  hopping in disordered semiconducting polymers: How accurate is the
  Miller--Abrahams model?}},\ }\href {https://doi.org/10.1063/1.3474618}
  {\bibfield  {journal} {\bibinfo  {journal} {Appl. Phys. Lett.}\ }\textbf
  {\bibinfo {volume} {97}},\ \bibinfo {pages} {043305} (\bibinfo {year}
  {2010})}\BibitemShut {NoStop}%
\bibitem [{\citenamefont {Filoche}\ and\ \citenamefont
  {Mayboroda}(2012)}]{FilocheUniversalMechanismAnderson2012}%
  \BibitemOpen
  \bibfield  {author} {\bibinfo {author} {\bibfnamefont {M.}~\bibnamefont
  {Filoche}}\ and\ \bibinfo {author} {\bibfnamefont {S.}~\bibnamefont
  {Mayboroda}},\ }\bibfield  {title} {\bibinfo {title} {Universal mechanism for
  {{Anderson}} and weak localization},\ }\href
  {https://doi.org/10.1073/pnas.1120432109} {\bibfield  {journal} {\bibinfo
  {journal} {Proc. Natl Acad. Sci. USA}\ }\textbf {\bibinfo {volume} {109}},\
  \bibinfo {pages} {14761} (\bibinfo {year} {2012})}\BibitemShut {NoStop}%
\bibitem [{\citenamefont {Arnold}\ \emph {et~al.}(2016)\citenamefont {Arnold},
  \citenamefont {David}, \citenamefont {Jerison}, \citenamefont {Mayboroda},\
  and\ \citenamefont {Filoche}}]{ArnoldEffectiveConfiningPotential2016}%
  \BibitemOpen
  \bibfield  {author} {\bibinfo {author} {\bibfnamefont {D.~N.}\ \bibnamefont
  {Arnold}}, \bibinfo {author} {\bibfnamefont {G.}~\bibnamefont {David}},
  \bibinfo {author} {\bibfnamefont {D.}~\bibnamefont {Jerison}}, \bibinfo
  {author} {\bibfnamefont {S.}~\bibnamefont {Mayboroda}},\ and\ \bibinfo
  {author} {\bibfnamefont {M.}~\bibnamefont {Filoche}},\ }\bibfield  {title}
  {\bibinfo {title} {Effective confining potential of quantum states in
  disordered media},\ }\href {https://doi.org/10.1103/PhysRevLett.116.056602}
  {\bibfield  {journal} {\bibinfo  {journal} {Phys. Rev. Lett.}\ }\textbf
  {\bibinfo {volume} {116}},\ \bibinfo {pages} {056602} (\bibinfo {year}
  {2016})}\BibitemShut {NoStop}%
\bibitem [{\citenamefont {Baranovskii}\ and\ \citenamefont
  {Rubel}(2017)}]{BaranovskiiChargeTransportDisordered2017}%
  \BibitemOpen
  \bibfield  {author} {\bibinfo {author} {\bibfnamefont {S.}~\bibnamefont
  {Baranovskii}}\ and\ \bibinfo {author} {\bibfnamefont {O.}~\bibnamefont
  {Rubel}},\ }\bibinfo {title} {Charge transport in disordered materials},\ in\
  \href {https://doi.org/10.1007/978-3-319-48933-9_9} {\emph {\bibinfo
  {booktitle} {Springer Handbook of Electronic and Photonic Materials}}},\
  \bibinfo {editor} {edited by\ \bibinfo {editor} {\bibfnamefont
  {S.}~\bibnamefont {Kasap}}\ and\ \bibinfo {editor} {\bibfnamefont
  {P.}~\bibnamefont {Capper}}}\ (\bibinfo  {publisher} {Springer International
  Publishing},\ \bibinfo {address} {Cham},\ \bibinfo {year} {2017})\ pp.\
  \bibinfo {pages} {1--1}\BibitemShut {NoStop}%
\bibitem [{\citenamefont {Oelerich}\ \emph {et~al.}(2012)\citenamefont
  {Oelerich}, \citenamefont {Huemmer},\ and\ \citenamefont
  {Baranovskii}}]{OelerichHowFindOut2012}%
  \BibitemOpen
  \bibfield  {author} {\bibinfo {author} {\bibfnamefont {J.~O.}\ \bibnamefont
  {Oelerich}}, \bibinfo {author} {\bibfnamefont {D.}~\bibnamefont {Huemmer}},\
  and\ \bibinfo {author} {\bibfnamefont {S.~D.}\ \bibnamefont {Baranovskii}},\
  }\bibfield  {title} {\bibinfo {title} {How to find out the density of states
  in disordered organic semiconductors},\ }\href
  {https://doi.org/10.1103/PhysRevLett.108.226403} {\bibfield  {journal}
  {\bibinfo  {journal} {Phys. Rev. Lett.}\ }\textbf {\bibinfo {volume} {108}},\
  \bibinfo {pages} {226403} (\bibinfo {year} {2012})}\BibitemShut {NoStop}%
\bibitem [{\citenamefont {Filoche}\ \emph {et~al.}(2017)\citenamefont
  {Filoche}, \citenamefont {Piccardo}, \citenamefont {Wu}, \citenamefont {Li},
  \citenamefont {Weisbuch},\ and\ \citenamefont
  {Mayboroda}}]{FilocheLocalizationLandscapeTheory2017}%
  \BibitemOpen
  \bibfield  {author} {\bibinfo {author} {\bibfnamefont {M.}~\bibnamefont
  {Filoche}}, \bibinfo {author} {\bibfnamefont {M.}~\bibnamefont {Piccardo}},
  \bibinfo {author} {\bibfnamefont {Y.-R.}\ \bibnamefont {Wu}}, \bibinfo
  {author} {\bibfnamefont {C.-K.}\ \bibnamefont {Li}}, \bibinfo {author}
  {\bibfnamefont {C.}~\bibnamefont {Weisbuch}},\ and\ \bibinfo {author}
  {\bibfnamefont {S.}~\bibnamefont {Mayboroda}},\ }\bibfield  {title} {\bibinfo
  {title} {Localization landscape theory of disorder in semiconductors. {I}.
  {T}heory and modeling},\ }\href {https://doi.org/10.1103/PhysRevB.95.144204}
  {\bibfield  {journal} {\bibinfo  {journal} {Phys. Rev. B}\ }\textbf {\bibinfo
  {volume} {95}},\ \bibinfo {pages} {144204} (\bibinfo {year}
  {2017})}\BibitemShut {NoStop}%
\bibitem [{\citenamefont {Arnold}\ \emph {et~al.}(2019)\citenamefont {Arnold},
  \citenamefont {David}, \citenamefont {Filoche}, \citenamefont {Jerison},\
  and\ \citenamefont {Mayboroda}}]{ArnoldComputingSpectraSolving2019}%
  \BibitemOpen
  \bibfield  {author} {\bibinfo {author} {\bibfnamefont {D.~N.}\ \bibnamefont
  {Arnold}}, \bibinfo {author} {\bibfnamefont {G.}~\bibnamefont {David}},
  \bibinfo {author} {\bibfnamefont {M.}~\bibnamefont {Filoche}}, \bibinfo
  {author} {\bibfnamefont {D.}~\bibnamefont {Jerison}},\ and\ \bibinfo {author}
  {\bibfnamefont {S.}~\bibnamefont {Mayboroda}},\ }\bibfield  {title} {\bibinfo
  {title} {Computing spectra without solving eigenvalue problems},\ }\href
  {https://doi.org/10.1137/17M1156721} {\bibfield  {journal} {\bibinfo
  {journal} {SIAM J. Sci. Comput.}\ }\textbf {\bibinfo {volume} {41}},\
  \bibinfo {pages} {B69} (\bibinfo {year} {2019})}\BibitemShut {NoStop}%
\bibitem [{\citenamefont {Hislop}\ and\ \citenamefont
  {Sigal}()}]{HislopIntroductionSpectralTheory1996}%
  \BibitemOpen
  \bibfield  {author} {\bibinfo {author} {\bibfnamefont {P.~D.}\ \bibnamefont
  {Hislop}}\ and\ \bibinfo {author} {\bibfnamefont {I.~M.}\ \bibnamefont
  {Sigal}},\ }\href {https://doi.org/10.1007/978-1-4612-0741-2} {\emph
  {\bibinfo {title} {Introduction to {{Spectral Theory}}: {{With Applications}}
  to {{Schr{\"o}dinger Operators}}}}},\ Applied {{Mathematical Sciences}}\
  (\bibinfo  {publisher} {{Springer-Verlag}})\BibitemShut {NoStop}%
\bibitem [{\citenamefont {Li}\ \emph {et~al.}(2017)\citenamefont {Li},
  \citenamefont {Piccardo}, \citenamefont {Lu}, \citenamefont {Mayboroda},
  \citenamefont {Martinelli}, \citenamefont {Peretti}, \citenamefont {Speck},
  \citenamefont {Weisbuch}, \citenamefont {Filoche},\ and\ \citenamefont
  {Wu}}]{LiLocalizationLandscapeTheory2017a}%
  \BibitemOpen
  \bibfield  {author} {\bibinfo {author} {\bibfnamefont {C.-K.}\ \bibnamefont
  {Li}}, \bibinfo {author} {\bibfnamefont {M.}~\bibnamefont {Piccardo}},
  \bibinfo {author} {\bibfnamefont {L.-S.}\ \bibnamefont {Lu}}, \bibinfo
  {author} {\bibfnamefont {S.}~\bibnamefont {Mayboroda}}, \bibinfo {author}
  {\bibfnamefont {L.}~\bibnamefont {Martinelli}}, \bibinfo {author}
  {\bibfnamefont {J.}~\bibnamefont {Peretti}}, \bibinfo {author} {\bibfnamefont
  {J.~S.}\ \bibnamefont {Speck}}, \bibinfo {author} {\bibfnamefont
  {C.}~\bibnamefont {Weisbuch}}, \bibinfo {author} {\bibfnamefont
  {M.}~\bibnamefont {Filoche}},\ and\ \bibinfo {author} {\bibfnamefont {Y.-R.}\
  \bibnamefont {Wu}},\ }\bibfield  {title} {\bibinfo {title} {Localization
  landscape theory of disorder in semiconductors. {III}. {A}pplication to
  carrier transport and recombination in light emitting diodes},\ }\href
  {https://doi.org/10.1103/PhysRevB.95.144206} {\bibfield  {journal} {\bibinfo
  {journal} {Phys. Rev. B}\ }\textbf {\bibinfo {volume} {95}},\ \bibinfo
  {pages} {144206} (\bibinfo {year} {2017})}\BibitemShut {NoStop}%
\bibitem [{\citenamefont {Logg}\ \emph {et~al.}(2012)\citenamefont {Logg},
  \citenamefont {Mardal},\ and\ \citenamefont
  {Wells}}]{LoggAutomatedSolutionDifferential2012}%
  \BibitemOpen
  \bibinfo {editor} {\bibfnamefont {A.}~\bibnamefont {Logg}}, \bibinfo {editor}
  {\bibfnamefont {K.-A.}\ \bibnamefont {Mardal}},\ and\ \bibinfo {editor}
  {\bibfnamefont {G.}~\bibnamefont {Wells}},\ eds.,\ \href
  {https://doi.org/10.1007/978-3-642-23099-8} {\emph {\bibinfo {title}
  {Automated {{Solution}} of {{Differential Equations}} by the {{Finite Element
  Method}}: {{The FEniCS Book}}}}},\ Lecture {{Notes}} in {{Computational
  Science}} and {{Engineering}}\ (\bibinfo  {publisher} {{Springer-Verlag}},\
  \bibinfo {year} {2012})\BibitemShut {NoStop}%
\bibitem [{\citenamefont {Hernandez}\ \emph {et~al.}(2005)\citenamefont
  {Hernandez}, \citenamefont {Roman},\ and\ \citenamefont
  {Vidal}}]{HernandezSLEPcScalableFlexible2005}%
  \BibitemOpen
  \bibfield  {author} {\bibinfo {author} {\bibfnamefont {V.}~\bibnamefont
  {Hernandez}}, \bibinfo {author} {\bibfnamefont {J.~E.}\ \bibnamefont
  {Roman}},\ and\ \bibinfo {author} {\bibfnamefont {V.}~\bibnamefont {Vidal}},\
  }\bibfield  {title} {\bibinfo {title} {{SLEPc: A Scalable and Flexible
  Toolkit for the Solution of Eigenvalue Problems}},\ }\href
  {https://doi.org/10.1145/1089014.1089019} {\bibfield  {journal} {\bibinfo
  {journal} {ACM Trans. Math. Softw.}\ }\textbf {\bibinfo {volume} {31}},\
  \bibinfo {pages} {351} (\bibinfo {year} {2005})}\BibitemShut {NoStop}%
\bibitem [{\citenamefont {Sethian}(1996)}]{SethianFastMarchingLevel1996}%
  \BibitemOpen
  \bibfield  {author} {\bibinfo {author} {\bibfnamefont {J.~A.}\ \bibnamefont
  {Sethian}},\ }\bibfield  {title} {\bibinfo {title} {A fast marching level set
  method for monotonically advancing fronts},\ }\href
  {https://doi.org/10.1073/pnas.93.4.1591} {\bibfield  {journal} {\bibinfo
  {journal} {Proc. Natl Acad. Sci. USA}\ }\textbf {\bibinfo {volume} {93}},\
  \bibinfo {pages} {1591} (\bibinfo {year} {1996})}\BibitemShut {NoStop}%
\bibitem [{\citenamefont {Kimmel}\ and\ \citenamefont
  {Sethian}(1998)}]{KimmelComputingGeodesicPaths1998}%
  \BibitemOpen
  \bibfield  {author} {\bibinfo {author} {\bibfnamefont {R.}~\bibnamefont
  {Kimmel}}\ and\ \bibinfo {author} {\bibfnamefont {J.~A.}\ \bibnamefont
  {Sethian}},\ }\bibfield  {title} {\bibinfo {title} {Computing geodesic paths
  on manifolds},\ }\href {https://doi.org/10.1073/pnas.95.15.8431} {\bibfield
  {journal} {\bibinfo  {journal} {Proc. Natl Acad. Sci. USA}\ }\textbf
  {\bibinfo {volume} {95}},\ \bibinfo {pages} {8431} (\bibinfo {year}
  {1998})}\BibitemShut {NoStop}%
\bibitem [{\citenamefont {Ambegaokar}\ \emph {et~al.}(1971)\citenamefont
  {Ambegaokar}, \citenamefont {Halperin},\ and\ \citenamefont
  {Langer}}]{AmbegaokarHoppingConductivityDisordered1971}%
  \BibitemOpen
  \bibfield  {author} {\bibinfo {author} {\bibfnamefont {V.}~\bibnamefont
  {Ambegaokar}}, \bibinfo {author} {\bibfnamefont {B.~I.}\ \bibnamefont
  {Halperin}},\ and\ \bibinfo {author} {\bibfnamefont {J.~S.}\ \bibnamefont
  {Langer}},\ }\bibfield  {title} {\bibinfo {title} {Hopping conductivity in
  disordered systems},\ }\href {https://doi.org/10.1103/PhysRevB.4.2612}
  {\bibfield  {journal} {\bibinfo  {journal} {Phys. Rev. B}\ }\textbf {\bibinfo
  {volume} {4}},\ \bibinfo {pages} {2612} (\bibinfo {year} {1971})}\BibitemShut
  {NoStop}%
\end{thebibliography}%

\end{document}


\section{Supplemental Material: Hopping Rates}

\begin{figure}[H]
 \centering
    \subfloat[\label{fig:rates_MA}]{
    	\includegraphics[width=0.48\columnwidth]{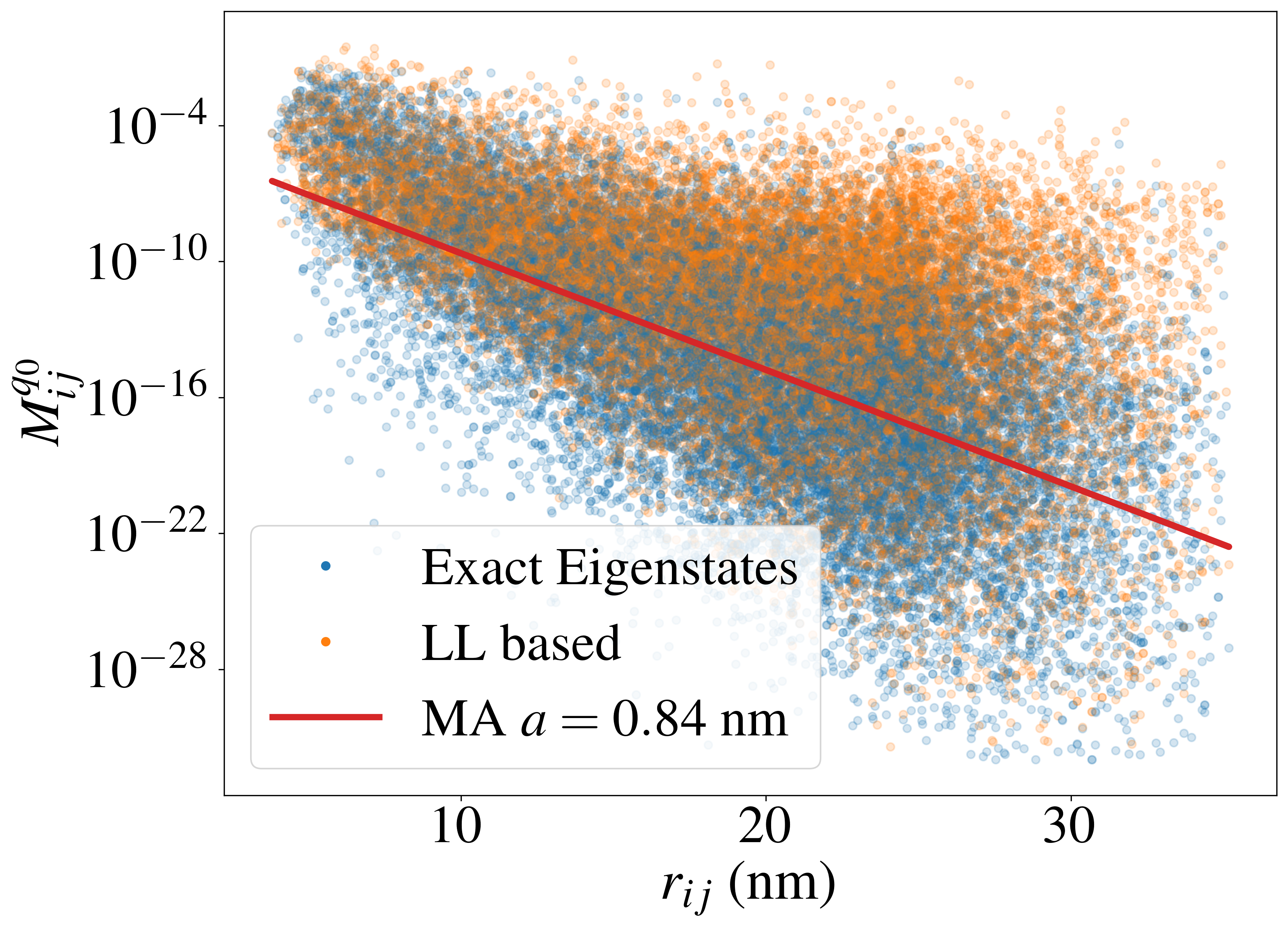}
    }\hfill
   	\subfloat[\label{fig:rates}]{
		\includegraphics[width=0.48\columnwidth]{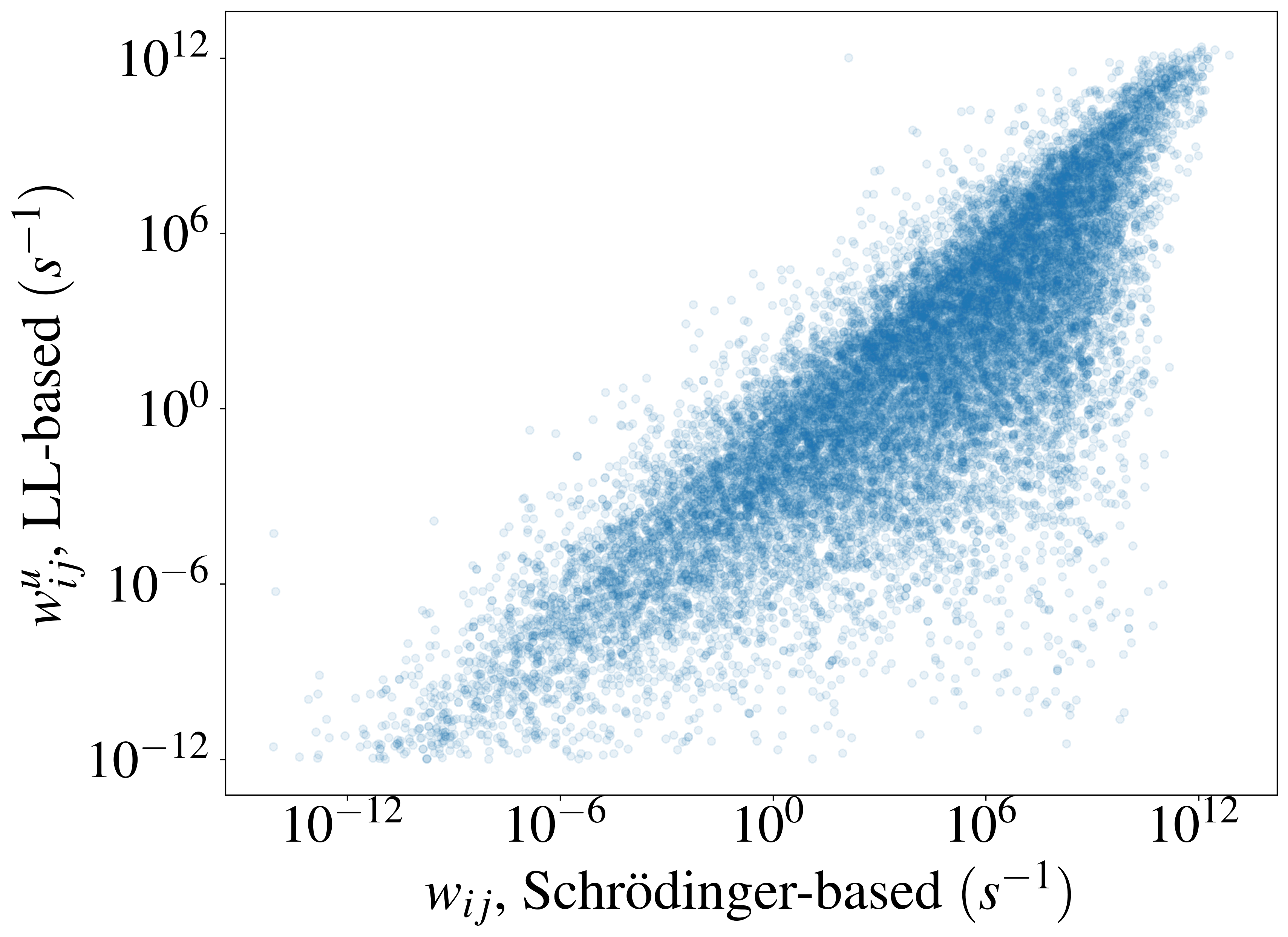}
	}
    \caption{a) Scatter plots of $M^{q_0}_{ij}$, the spatial component of the hopping rates (see Eq.~\ref{eq:spatial}) computed with the exact eigenstates (blue dots) and with the LL-based states (orange dots), as a function of the distance between the states. In red, we display the exponential fit used in the Miller-Abrahams model, $M^{q_0}_{ij} = \exp( -r_{ij}/a)$. (b) Comparison between the hopping rates based on the exact eigenstates $w_{ij}$ (horizontal axis) and the LL-based states $w^{u}_{ij}$ (vertical axis) for 50~realizations of the disordered potential (see description in the body of the paper).
    }
   \label{fig:rates_comp}
\end{figure}

In Fig.~\ref{fig:rates_comp}, we compare the hopping rates $w_{ij}$ computed (i) from the exact eigenstates, (ii) from the localization landscape (LL) theory, and (iii) from the Miller-Abrahams model. The hopping rates are defined by 
\begin{equation}
w_{ij} = \frac{D^2q_0^3}{8\pi^2\rho_m\hbar c_s^2} \, \abs{M^{q_0}_{ij}}^2 \, \Big\{ n_B + \frac{1}{2} \pm \frac{1}{2}\Big\} \,,
\label{eq:hopping_3d}
\end{equation}
where $D$ is the deformation potential constant, $\rho_m$ is the mass density of the material, $c_s$ is the speed of sound in the material, $n_B(E_q, T)$ is the average occupation number of a phonon with energy $E_q$ at temperature $T$ (given by the Bose-Einstein statistics), $q_0 = |E_j - E_i|/\hbar c_s$ and $M^{q_0}_{ij}$ is given by the following overlap integral:
\begin{equation}
M^{q_0}_{ij} = \int_{q = q_0} d\Omega_q \int d\vb{r} ~e^{-i\vb{q} \cdot \vb{r}}~\psi^{*}_i(\vb{r}) \psi_j(\vb{r}) \,.
\label{eq:spatial}
\end{equation}
The LL-based states $\psi^{(u)}_i$ are defined as
\begin{equation}
    \psi^{(u)}_i(\vb{r}) = c_i
    \begin{dcases}
        E^*_i \, u(\vb{r}) & \text{inside}~B_i\\
        \exp(-\rho_{E^*_i}( \vb{r}, B_i)) &  \text{outside}~B_i
    \end{dcases}
\end{equation}
where $c_i$ is a normalization constant, the basin $B_i$ being defined as the connected domain around the local minimum of $1/u$ whose boundary is the level set $u(\vb{r}) = 1/E^*_i$. We use material parameters similar to those of disordered alloys of InGaN: $D = \unit{8.3}{eV}$, $c_s = \unit{8 \cdot 10^3}{m \cdot s^{-1}}$ and $\rho_m = \unit{6150}{kg \cdot m^{-3}}$.

%